\documentclass[aps, reprint,superscriptaddress,amsmath,amssymb, twocolumn]{revtex4-2}
\pdfoutput=1

\usepackage[utf8]{inputenc}

\usepackage{hyperref}
\usepackage{pgfplots}
\usepackage{enumitem}
\usepackage{physics}
\usepackage{float}
\usepackage{bbold}

\usepackage{graphicx}

\begin{document}
	
	\title{Ambiguous Resonances in Multipulse Quantum Sensing \\ with Nitrogen Vacancy Centers}
	
	\author{Lucas~Tsunaki}
	
	\affiliation{Department Spins in Energy Conversion and Quantum Information Science (ASPIN), Helmholtz-Zentrum Berlin für Materialien und Energie GmbH, Hahn-Meitner-Platz 1, 14109 Berlin, Germany}
	
	\author{Anmol~Singh}
	\affiliation{Department Spins in Energy Conversion and Quantum Information Science (ASPIN), Helmholtz-Zentrum Berlin für Materialien und Energie GmbH, Hahn-Meitner-Platz 1, 14109 Berlin, Germany}
	
	\author{Kseniia~Volkova}
	\affiliation{Department Spins in Energy Conversion and Quantum Information Science (ASPIN), Helmholtz-Zentrum Berlin für Materialien und Energie GmbH, Hahn-Meitner-Platz 1, 14109 Berlin, Germany}
	
	\author{Sergei~Trofimov}
	\affiliation{Department Spins in Energy Conversion and Quantum Information Science (ASPIN), Helmholtz-Zentrum Berlin für Materialien und Energie GmbH, Hahn-Meitner-Platz 1, 14109 Berlin, Germany}
	
	\author{Tommaso~Pregnolato}
	\affiliation{Department of Physics, Humboldt-Universität zu Berlin, Newtonstraße 15, 12489 Berlin, Germany}
	\affiliation{Ferdinand-Braun-Institut (FBH), Gustav-Kirchhoff-Straße 4, 12489 Berlin, Germany}
	
	\author{Tim~Schröder}
	\affiliation{Department of Physics, Humboldt-Universität zu Berlin, Newtonstraße 15, 12489 Berlin, Germany}
	\affiliation{Ferdinand-Braun-Institut (FBH), Gustav-Kirchhoff-Straße 4, 12489 Berlin, Germany}
	
	\author{Boris~Naydenov}
	\email{boris.naydenov@helmholtz-berlin.de}	
	\affiliation{Department Spins in Energy Conversion and Quantum Information Science (ASPIN), Helmholtz-Zentrum Berlin für Materialien und Energie GmbH, Hahn-Meitner-Platz 1, 14109 Berlin, Germany}
	\affiliation{Berlin Joint EPR Laboratory, Fachbereich Physik, Freie Universität Berlin, 14195 Berlin, Germany}
	
	\begin{abstract}
		
		Dynamical decoupling multipulse sequences can be applied to solid state spins for sensing weak oscillating fields from nearby single nuclear spins. By periodically reversing the probing system's evolution, other noises are counteracted and filtered out over the total evolution. However, the technique is subject to intricate interactions resulting in additional resonant responses, which can be misinterpreted with the actual signal intended to be measured. We experimentally characterized three of these effects present in single nitrogen vacancy centers in diamond, where we also developed a numerical simulations model without rotating wave approximation, showing robust correlation to the experimental data. Regarding centers with the $^{15}$N nitrogen isotope, we observed that a small misalignment in the bias magnetic field causes the precession of the nitrogen nuclear spin to be sensed by the electronic spin of the center. Another studied case of ambiguous resonances comes from the coupling with lattice $^{13}$C nuclei, where we used the echo modulation frequencies to obtain the interaction Hamiltonian and then utilized the latter to simulate multipulse sequences. Finally, we also measured and simulated the effects from the free evolution of the quantum system during finite pulse durations. Due to the large data volume and the strong dependency of these ambiguous resonances with specific experimental parameters, we provide a simulations dataset with a user-friendly graphical interface, where users can compare simulations with their own experimental data for spectral disambiguation. Although focused with nitrogen vacancy centers and dynamical decoupling sequences, these results and the developed model can potentially be applied to other solid state spins and quantum sensing techniques.
		
	\end{abstract}
	
	\maketitle
	
	
\section{Introduction}\label{sec:intro}

The second quantum revolution is driving sensing technology \cite{quantum_sensing} to ever-increasing precision and accuracy, thus opening doors to novel measurement possibilities \cite{LIGO, cancer1, cancer2, biological_compatibility, single1, single2, single3, strong_coupling}. This quantum sensing advantage relies on the fact that quantum properties are extremely sensitive to their environment. However, such key element is also a vulnerability, as quantum sensors are susceptible to complex interactions with their environment, leading to responses that can be misinterpreted with the actual signal intended to be measured.

\begin{figure*}[t!]
	\includegraphics[width=\textwidth]{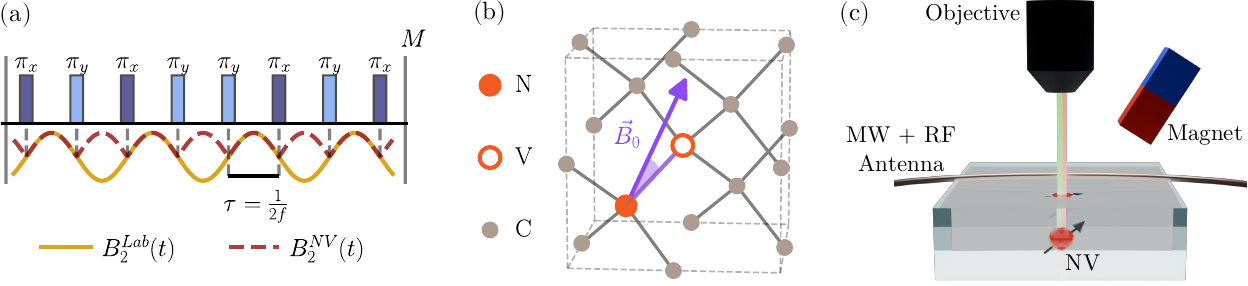}
	\caption{(a) The XY8-$M$ dynamical decoupling sequence is composed of eight alternated $\pi_{x,y}$-pulses repeated $M$ times, which counteracts oscillating magnetic field noises except for frequencies corresponding to half of the pulse separation $f_0=1/(2\tau)$ and odd multiples of it. This way, it can be applied with a probing spin to sense weak oscillating fields from single nuclear spins. (b) The NV center is composed of a nitrogen substitution adjacent to a vacancy in the carbon lattice of diamonds. This results in an electronic spin triplet or singlet configurations, with well-defined quantum states within the band-gap of diamond. In general, the external magnetic field $\Vec{B}_0$ forms an angle $\theta_0$ with the NV axis. \textsuperscript{13}C isotopes with spin 1/2 can also be present near the center. (c) Schematic representation of the experimental setup. An air objective focuses the green laser to polarize the NV in the $m_s=0$ state, which is also used to collect its red fluorescence and determine the $m_s$ state. The bias magnetic field $\Vec{B}_0$ is applied with a permanent magnet, while microwave pulses for coherent control of the electronic spin are transmitted through a wire on top of the diamond.}
	\label{fig1}
\end{figure*}

In this study, we consider these ambiguous resonant responses which affect multipulse dynamical decoupling (DD) pulse sequences. More specifically, the XY8-$M$ \cite{CPMG, DD_boris} and other similar sequences, as schematically represented in Fig. \ref{fig1} (a). 
By periodically reversing the quantum evolution of the system with intercalated $\pi_x$ and $\pi_y$ pulses, dephasings from the environment are counteracted over the total evolution. Thus, making the sequence into a narrow filter for frequencies corresponding to double of the pulse separation $f_0=1/(2\tau)$ and odd multiples of it \cite{multipulse1, spurious}: $3f_0$, $5f_0$, $7f_0$. Like so, weak signals from single nuclear spins can be filtered from much stronger background fields even at room temperature \cite{single1, single2, single3}, thus representing a remarkable improvement compared to conventional NMR, which requires large nuclear ensembles and much stronger bias magnetic fields. This improvement comes at the expense that the probing system is now intrinsically non-classical, hence having knowledge of its time evolution is indispensable for the  signal analysis. To overcome these challenges, an in-depth characterization of such ambiguous resonances experienced by the quantum sensor is necessary for the continued development of the field.

Despite being applicable to two-level systems in general, DD techniques excel with color center defects in solids \cite{color_centers}, given their versatility to sense different physical quantities and relative ease of experimental realization. Among them, the nitrogen vacancy (NV) center in diamond has been an epicenter of research due to its remarkable properties at a wide range of temperatures and compatibility with biological systems \cite{NV_review1, NV_review2}, unlike for instance, superconducting quantum interference devices (SQUIDs) \cite{SQUID}. NV centers are a substitutional defects in the diamond lattice where a carbon atom is replaced by nitrogen $^{14}$N ($I^n=1$) or $^{15}$N ($I^n=1/2$) adjacent to a vacancy [Fig. \ref{fig1} (b)]. The NV$^{-}$, denoted as NV for simplicity, forms a spin triplet configuration ($S=1$), with well-defined quantum states within the bandgap of diamond \cite{NV_configuration}. They can be efficiently polarized to the $m_s=0$ state with a non-resonant green laser excitation via an optical pumping mechanism \cite{optical_pumping}, which can also be used to read the population difference between $m_s=0$ and $m_s=\pm 1$ based on the resulting fluorescence. Finally, coherent transitions between the $m_s$ states can be achieved by applying resonant microwave (MW) pulses with controllable phase and duration \cite{rabi}. This way, a DD sequence consists of three parts: laser initialization, MW manipulation and fluorescence readout. For NVs, $\pi$/2-pulses are included before and after the sequence, in order to drive the electronic spin to and from the plane perpendicular to the quantization axis.

In the context of single NV multipulse sequences, ambiguous resonances can arise from different factors. We begin with Sec. \ref{sec:theory} giving a general theoretical framework used in our simulations, following up with a description of the experimental and simulation methods in Sec. \ref{sec:methods}. We then discuss the various ambiguous resonances appearing due to either external magnetic field misalignment and hyperfine interaction with $^{15}$N nuclei (Sec. \ref{sec:15N}), or coupling with lattice $^{13}$C having $I^c=1/2$ (Sec. \ref{sec:13C}), or free evolution of the system during finite pulses duration \cite{spurious} (Sec. \ref{sec:pulse_duration}). In sight of the numerous possible ambiguous resonances specific to experimental parameters, we conclude with Sec. \ref{sec:discussion} providing a dataset with graphical interface where simulations can be visualized and compared with the user's experimental data for spectral disambiguation of resonances. Although the discussion is focused on NVs, some of these effects are also relevant to other color centers and quantum systems in general.

	
\section{Theoretical Framework}\label{sec:theory}

In contrast to conventional NMR, in DD quantum sensing, the probing system is an inseparable part from the target spin and the exact Hamiltonian of the quantum sensor needs to be taken into account. The total Hamiltonian of a NV in a multipulse sequence can be decomposed in three parts as $H = H_0 + H_1(t) + H_2(t)$, where $H_0$ is the internal component, $H_1(t)$ is the control Hamiltonian describing the interaction with the MW pulses and $H_2(t)$ the sensing interaction itself. For the first internal component, we consider a spin model extensively studied and characterized \cite{NV_hamiltonian3, NV_hamiltonian1, NV_hamiltonian2}. It can be expressed in the Hilbert space of the electronic and nitrogen nuclear spins $\mathcal{H}_S \otimes \mathcal{H}_{I^n}$ in factors of decreasing energy as
\begin{alignat}{2}\label{eq:1}
	H_0 = D\left( {S_z}^2 -\frac{S^2}{3} \right) \tag{1.1} \\
	- \gamma^e \Vec{B}_0 \cdot \Vec{S} &\tag{1.2} \\
	+ a_{\perp}^n \left( S_x I_x^n + S_y I_y^n \right) + a_{\parallel}^n S_z I_z^n &\tag{1.3} \\
	-  \gamma^n \Vec{B}_0 \cdot \Vec{I}^n , \tag{1.4}&
\end{alignat}\setcounter{equation}{1}
where the $z$ axis is taken along the NV axis. The first line (Eq. 1.1) corresponds to the zero-field splitting due to the dipolar interaction between the two electrons, where $D=2.87$~GHz is the axial component. The second and fourth lines (Eqs. 1.2 and 1.4) correspond to Zeeman interactions with the external magnetic field $\Vec{B_0}$ for the electronic spin and nitrogen nuclear spin respectively, with gyromagnetic ratios $\gamma^e=-28025$~MHz/T, $\gamma^n = -4.316$~MHz/T for $^{15}$N and $\gamma^n = 3.077$~MHz/T for \textsuperscript{14}N. Here we consider a general case where $\Vec{B}_0$ is not necessarily aligned with the NV axis, as usually assumed. The third line (Eq. 1.3) is the hyperfine interaction between spins which is diagonal in this symmetry reference, with $a_\perp^n=-2.70$~MHz and $a_{\parallel}^n=-2.14$~MHz for $^{14}$N, while $a_\perp^n=3.65$~MHz and $a_{\parallel}^n=3.03$~MHz for $^{15}$N. When considering \textsuperscript{14}N, there is an extra quadrupole term given by $Q (I_z^n) ^2$, where $Q=-5.01$~MHz. Although small, the presence of the quadrupole interaction fundamentally changes the dynamics of the system, as the interaction tends to fix the nuclear spin orientation and suppress the Larmor precession of $^{14}$N nuclear spins for magnetic fields $B_0\ll Q/\gamma^n = 1.63$ T \cite{13C_1}.

The control Hamiltonian $H_1(t)$ for an oscillating field along a vector $\Vec{B}_1$, with angular frequency $\Omega_{MW}$ and phase $\phi_1$ can be modeled as
\begin{equation}\label{eq:2}
	H_1(t) = \gamma^e  \sin(\Omega_{MW} t + \phi_1) \Vec{B}_1 \cdot \Vec{S}.
\end{equation}
Typically, a rotating wave approximation (RWA) is assumed \cite{slichter}. Thus, by setting the pulse frequency in resonance with one of the $m_s=0 \leftrightarrow \pm1$ transitions, $H_1(t)$ can be treated as a time independent operator which induces a rotation on the Bloch sphere of the respective manifold, where the pulse duration controls the angle and $\phi_1$ the axis of rotation in the rotating frame. Although this approximation often results in accurate descriptions of the electronic states transitions, it can overlook important interactions which lead to the ambiguous resonant responses. In particular, the $S_{x,y}$ components of the electronic Zeeman and the perpendicular terms of the hyperfine interactions $a_{\perp}^n \left( S_x I_x^n + S_y I_y^n \right)$ do not commute with the rotation operator and result in extra resonances (Sec. \ref{sec:15N}). This time dependency resulting from the non-commuting terms could still be handled as a perturbation to the secular part of the Hamiltonian up to 2\textsuperscript{nd} order of the dominant term $(1/D)^p$ \cite{13C_1, RWA_1, RWA_2}. However, this approach is also limited, as it overlooks the possibility of double quantum transitions caused by multi-frequency or broad band excitations \cite{double_quantum}, as well as it implies small bias fields $|\gamma_e \Vec{B}_0|\ll D$ and weak driving regimes $\gamma_e B_1 \ll D$, which is not always experimentally desired due to extended pulse durations \cite{strong_driving_2}. Furthermore, perturbation theory, in such circumstances, restricts on calculating approximated eigenvalues for the perturbed Hamiltonian, ignoring exact characteristics of the MW field and the free-evolution of the system during finite pulse durations. Which also leads to other ambiguous resonances (Sec. \ref{sec:pulse_duration}) \cite{spurious}.

For a single nuclear spin in the strong coupling regime but without contact term \cite{strong_coupling}, the sensing Hamiltonian $H_2$ can be simply expressed in an extended Hilbert space to include the sensed target spin $\mathcal{H}_S \otimes \mathcal{H}_{I^n} \otimes \mathcal{H}_{I^c}$, where we denote the sensed spin operator $I^{c}$ as carbon nuclei for a concrete example. For a spin-1/2 and neglecting the interaction between the nitrogen and target nuclear spin, this Hamiltonian is given by a hyperfine coupling between the electronic spin operator and the sensed nuclear spin added to the Zeeman interaction with the external magnetic field
\begin{equation}\label{eq:3}
	H_2 = \Vec{S} \cdot A^{c} \cdot \Vec{I}^c - \gamma^c \Vec{B}_0 \cdot \Vec{I}^c .
\end{equation}
Diagonalizing the total Hamiltonian $H_0+H_2$ results in distinct target nuclear spin Larmor frequencies $\tilde{\omega}_{m_s}^c$, dependent on the electronic spin state. Given that the hyperfine matrix $A^{c}$ must be symmetrical, the sensing interaction is completely described by its six components plus the gyromagnetic ratio $\gamma^{c}$. The same Hamiltonian can also be used to describe an ensemble of nuclear spins, provided that their states are represented by density operators. Apart from nuclear spins, it is also desirable to sense oscillating classical fields. Particularly for this study, an externally applied RF signal has the advantage of having controllable intensity and frequency. In this case, the interaction is described in terms of a field $\Vec{B}_2(t)$ with frequency $f_0$ as
\begin{equation}\label{eq:4}
H_2(t) = \gamma^e \sin(2 \pi f_0 t + \phi_2) \Vec{B}_2 \cdot \Vec{S} ,
\end{equation}
in analogy to Eq. \ref{eq:2}.

In a simplified manner, an ensemble of nuclear spins could also be thought of in terms of a semi-classical oscillating field \cite{strong_coupling, phd_muller}. Despite their average magnetization being close to zero at room temperature, their collective precession around $\Vec{B}_0$ results in an effective oscillating magnetic field with statically varying amplitude and phase, but with frequencies centered around the effective nuclear Larmor frequencies $\tilde{\omega}_{m_s}^c$ at each electronic state $m_s$. On the other hand, a weak classical RF signal could emulate a precessing nuclear spin and be approximately described in terms of an effective hyperfine interaction. Hence, an analogy can be drawn between the quantum and semi-classical descriptions from Eqs. \ref{eq:3} and \ref{eq:4}, where the interaction is either represented in terms of the six components of the 3-dimensional symmetric hyperfine coupling matrix $A^c_{ij}$ plus the gyromagnetic ration $\gamma_s$, or the three nuclear precession frequencies $\tilde{\omega_{0,\pm1}^c}$ plus four parameters of the effective field vector $\phi_2$ and $\vec{B}_0(x,y,z)$. Strictly speaking, both Hamiltonians describe fundamentally different processes, yet they share mathematical similarities and can lead to resembling effects at low order XY8-$M$ sequences, as further shown in Appendix \ref{S5:pulse_errors}.

Altogether, a complete description of the pulsed dynamics of the NV subject to a multipulse sequence needs to account for the time-dependent MW pulses and oscillating RF field in the laboratory reference frame, while operating over mixed states. Within these considerations, the time evolution of the system can be solved by the Liouville-von Neumann equation
\begin{equation}\label{eq:5}
	i \hbar \dot{\rho}(t) = [H_0 + H_1(t) + H_2(t), \rho(t) ],
\end{equation}
with $H_1(t)$ being present only during MW pulses but $H_0$ and $H_2(t)$ during the entire time evolution. This results in a set of coupled differential equations for the elements in the density matrix representing the whole system $\rho(t)$ \cite{ME}, which can be integrated numerically in discretized small time steps \cite{qutip_1, qutip_2}. Another advantage of this method is that the non-unitary dynamics can be easily incorporated with the Lindblad equation \cite{lindblad}, using the associated collapse operators of decoherence and relaxation. Lastly, the observable measured after the DD sequence, i.e. the fluorescence, is given by the expectation value of the bright state density operator $ p = \Tr [\rho_0 \rho ^\dagger(t_f) ]$, where $\rho_0$ is the initial bright state and $\rho(t_f)$ the final state obtained from the equation. As discussed in Sec. \ref{sec:intro}, the electronic spin is optically initialized at $m_s=0$, while the nitrogen and target nuclear spins are in thermal equilibrium given by the Boltzmann distribution, which at room temperature leads to an identity density matrix. Thus, the initial state considered in the simulations is $\rho_0 \equiv \ket{0}\bra{0} \otimes \mathbb{1} \otimes \mathbb{1}$, where the dimensions of $\mathbb{1}$ dependent on the spin dimension of the nuclei.
	

\section{Experimental and Simulation Methods}\label{sec:methods}

Experimental data was obtained with a home-built confocal microscope as described in \cite{setup_1, setup_2} and schematically represented in Fig. \ref{fig1} (c). Single NVs are addressed with a green laser in continuous or pulsed modes, while their fluorescence is measured by a single photon detector. MW pulses are produced by an arbitrary waveform generator, which are then amplified and transmitted to the diamond by a thin copper wire. RF fields are generated continuously with controllable power and frequency, combined and then transmitted without amplification with the already amplified MW pulses. The external bias magnetic field $B_0$ is applied with a permanent magnet with four degrees of movement, while its intensity and the angle formed with the NV axis $\theta_0$ are calculated by the asymmetry between peaks in the continuous wave or pulsed optical detected magnetic resonance (ODMR) spectrum \cite{misalignement}. Lastly, the software package Qudi \cite{qudi} is used for hardware control and data acquisition.

In order to study each ambiguous resonance, different properties were required for the diamond samples. For measuring $^{13}$C ambiguous resonances, Sec. \ref{sec:13C}, a CVD grown sample with $^{14}$N was chosen in order to avoid $^{15}$N effects. Conversely, in Sec. \ref{sec:15N}, Hahn echo modulations and XY8-$M$ were measured in shallow implanted $^{15}$N diamonds. For longer XY8-12 measurements (Sec. \ref{sec:pulse_duration}), a deeper implanted $^{15}$N diamond sample was used with longer coherence times. A more detailed description of the experimental setup and diamond samples can be found in Appendix \ref{S1:exp} and \ref{S2:diamonds}.

The time evolution of the pulsed dynamics were calculated using Quantum Color Centers Analysis Toolbox (QuaCCAToo) \cite{quaccatoo} Python package, based on the master equation solver from Quantum Toolbox in Python (QuTiP) \cite{qutip_1, qutip_2}. Some of the most important simulations are open-source and provided in the Example Notebooks section of the QuCAAToo package. Experimentally measured fluorescence and simulated transition probability are compared through a linear relation, physically representing experimental uncertainties, as the collection efficiency or background light. The correlation $r$ between them is quantified from -1 to 1 by the Pearson coefficient \cite{pearson}, see Appendix \ref{S3:pearson} for more details. In Sec. \ref{sec:pulse_duration}, where the coupling parameters of $H_2$ are not defined, the transition probability was simulated over several points in parameter space and the simulation with the highest correlation coefficient was assumed.

\begin{figure*}[t!]
	\includegraphics[width=\columnwidth]{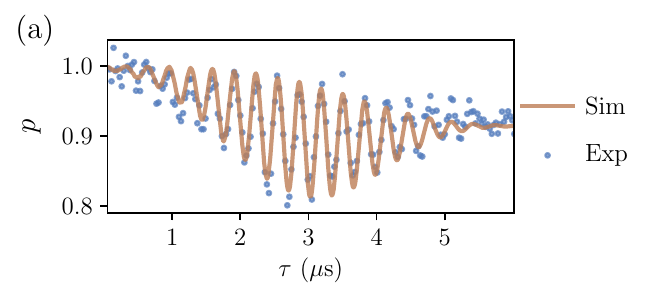}\hfill
	\includegraphics[width=\columnwidth]{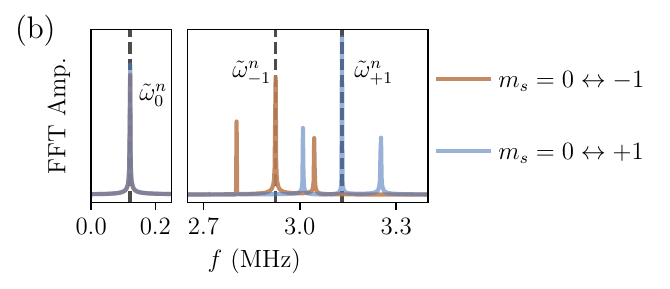}
	\includegraphics[width=\columnwidth]{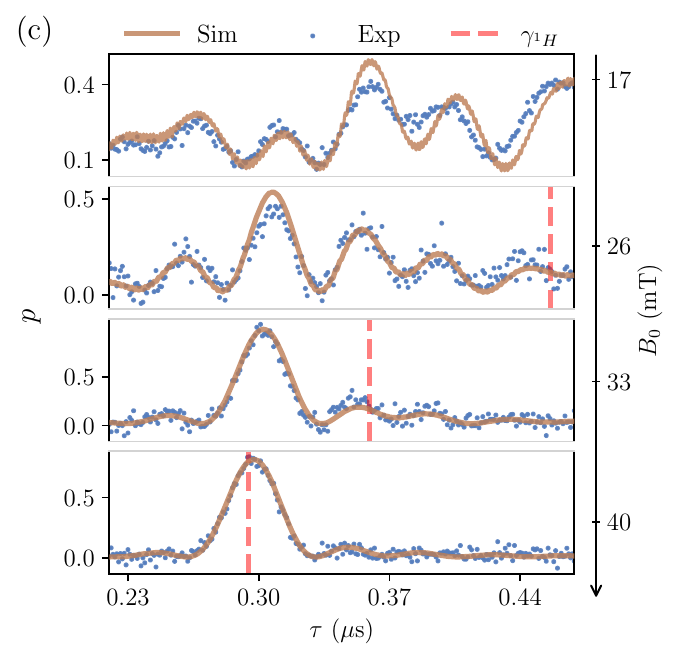}\hfill
	\includegraphics[width=\columnwidth]{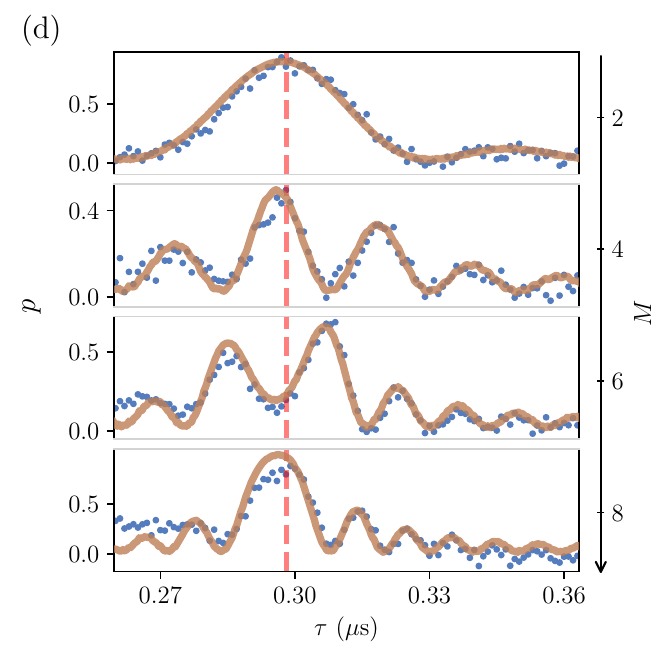}
	\caption{ (a) Experimental and simulated Hahn echo decay for $m_s=0\leftrightarrow + 1$ transition at $B_0=24$~mT and misalignment of $\theta_0=2.1^\circ$. The field misalignment leads the Larmor precession of the $^{15}$N nuclear spin to be sensed by the electronic spin, resulting in the envelope modulations \cite{phd_myers, 13C_1, RWA_1}. The coherence decay is fitted by $\exp[-(t/\tau_c)^4 ]$ \protect\cite{13C_1} with $\tau_c = 4.018\pm0.007$~$\mu$s and correlation $r=0.934$. The y-axis represents the simulated transition probability $p$, while the experimental points are given by the fluorescence counts linearly fitted with the simulation data (Sec. \ref{sec:methods}). This way, some of the experimental points appear above $p=1$ due to experimental noise, even though this would be mathematically impossible. (b) FFT of the simulated Hahn echo decay for both $m_s=0\leftrightarrow\pm 1$ transitions. The envelope modulations are defined by the effective Larmor frequencies of \textsuperscript{15}N $\omega_{m_s} ^n$, numerically simulated from the Hamiltonian. Less intense sidebands from $\omega_{\pm 1} ^n$ are also present in the spectra. (c) Experimental and simulated XY8-2 of the $^{15}$N resonances for different values of $B_0$ with $\theta_0$=(17.0, 9.3, 5.1, 2.7)$^\circ$ and correlations $r$=(0.858, 0.890, 0.971, 0.986), respectively. The resonance peaks have varying amplitude at almost constant pulse separations, evidencing a dominance of the hyperfine interaction with the electronic spin over the Zeeman interaction of the $^{15}$N spin with $B_0$. The expected resonances from $^1$H are also shown at $\tau_{H} = 1/(2 \gamma_H B_0)$. (d) XY8-$M$ at $B_0=39$~mT and $\theta_0=2.6^\circ$ with $r$=(0.975, 0.921, 0.945, 0.906), respectively. At this field, the ambiguous resonance could be mistakenly attributed to $^1$H nuclear spins. The high correlation between simulations and experimental data for longer pulse sequences further corroborate the numerical model.}
	\label{fig:15N}
\end{figure*}


\section{Results}

\subsection{Magnetic Field Misalignment and \textsuperscript{15}N Coupling}\label{sec:15N}

Perfectly aligning the bias field with the NV axis is typically only achievable using permanent magnets with precise rotating stages \cite{RWA_1}, or superconducting magnets. When considering manually controlled stages, as in our setup, the magnetic field alignment is limited to about $\Delta \theta_0 = 1^{\circ}$. In these circumstances, the terms $S_{x,y}$ in the Zeeman interaction (see Eq. \ref{eq:1}) and perpendicular nitrogen hyperfine terms can induce undesired electronic spin flips during multipulse sequences. This is better understood in another reference frame where $\Vec{B}_0$ is aligned along the new $z$ axis, causing the nitrogen hyperfine matrix to assume non-diagonal terms just like in the sensing Hamiltonian of Eq. \ref{eq:3}. Consequently, the misaligned field can cause the NV's electronic spin to sense its own nitrogen nuclear spin. Although this effect happens with both nitrogen isotopes, for $^{14}$NV it is nearly negligible at moderate fields due to the quadrupole interaction (Sec. \ref{sec:theory}), thus we restrict the discussion to $^{15}$NV in this section.

The influence of $^{15}$N coupling in simple pulsed experiments is already well-known \cite{phd_myers}, resulting in an Electron Spin Echo Envelope Modulation (ESEEM), as first observed by Rowan \textit{et al.} (1965) \cite{Mims65} in electron paramagnetic resonance experiments. However, these approaches are restricted to a semi-analytical framework \cite{13C_1, RWA_1} under certain approximations, as discussed in Sec. \ref{sec:theory}. Here we show that the measured Hahn echo decay also agrees with our numerical simulation model with a high correlation of $r=0.934$, Fig. \ref{fig:15N} (a). The electronic coherence decay presents envelope modulations determined by the effective nitrogen Larmor frequencies $\tilde{\omega}_{m_s}^n$, as shown by the fast Fourier transform (FFT) taken from the simulated data, Fig. \ref{fig:15N} (b). As the number of pulses increases however, an analytical approach for solving Eq. \ref{eq:5} and calculating this nitrogen coupling effect becomes increasingly complex.

To study these effects in longer DD sequences, XY8-2 for the $m_s=0\leftrightarrow+1$ transition was measured and simulated over a fixed $\tau$ range varying the magnetic field amplitude and angle, as shown in Fig. \ref{fig:15N} (c). The frequency of nitrogen-related resonances does not depend linearly on the applied magnetic field, as one would expect from a simple Zeeman relation. In contrast, the hyperfine interaction, dominating the energy levels of the nuclei, results in peaks with varying intensity and nearly constant position as the field changes. The peaks positions do not correspond to the effective Larmor frequencies of the nitrogen nuclei $\tilde{\omega}_{m_s}^n$ and a analytical model which describes their resonant frequency values remains an open problem. The lowest magnetic field shows the worst correlation between experiment and simulation of $r=0.858$, where a pulse duration error had to be included in the simulation. This causes a fast modulation of the signal, as discussed in details in Appendix \ref{S3:pearson}. This smaller correlation can be attributed to a larger misalignment angle of $\theta_0=17.0^\circ$, thus with a larger uncertainty.

These resonances differ for each $m_s=0\leftrightarrow \pm 1$ transition due to distinct nuclear state energies and are also present at longer $\tau$ values given by the odd multiples of their resonant frequencies, as previously observed for $^{13}$C resonances \cite{single1}. At long pulse separations of several $\mu$s, they mix and become a near continuum, which can be visualized in the simulations dataset. $^{15}$N resonances do not seem to occur at short pulse spacings of about less than 250 ns. Still, short $\tau$ are occupied by high frequency noises from paramagnetic impurities and even fractions of the resonant frequencies (Sec. \ref{sec:pulse_duration}) \cite{spurious}. Moreover, setting the resonant frequencies of target nuclear spins with low gyromagnetic ratios $\gamma^c$ at $\tau<250$~ns is technically challenging, as it requires large bias fields $B_0$, additionally with broad MW amplifiers for $\Omega_{MW}$ (Eq. \ref{eq:2}). Thus, these resonances from $^{15}$N are just unavoidable in many cases, underlying the necessity for a dataset where users can compare their experimental data for disambiguation. It is important to note, however, that not all $^{15}$NV centers under misaligned fields will present these ambiguous resonances, as additional interactions might dominate over it.

Fig. \ref{fig:15N} (c) also illustrates how these resonances could be mistakenly attributed to some other nuclei. The $^{15}$N resonance is compared with the expected Larmor frequency resonance of the commonly measured $^{1}$H nuclei $\tau_{H} = 1/(2 \gamma^h B_0)$, where $\gamma^h=42.57$~MHz/T. It can be seen that at 39 mT, the ambiguous resonance closely resembles a $^{1}$H nucleus. In Fig. \ref{fig:15N} (d), the signal dependence on the order of the pulse sequence $M$ is shown, while $B_0=39$~mT is kept constant. A strong agreement between the experimental data and the simulations is also observed for these longer sequences with many pulses, further corroborating the numerical model.


\subsection{\textsuperscript{13}C Coupling}\label{sec:13C}

Carbon nuclear isotope \textsuperscript{13}C with spin-1/2 is present in most diamonds with 1.1\% natural abundance. $^{13}$C sensing and manipulation by multipulse sequences is itself an active topic of research in sight of the potential for long-lived quantum memories \cite{single1, DD_13C_2} and also for Dynamic Nuclear Polarization applications (DNP) \cite{single_nv_dnp,dnp_review}, but in this case, it can act as another source of ambiguous resonances. Compared to $^{15}$N in the previous section, the Hamiltonian becomes more complex, as there are six free parameters from the hyperfine interaction $A^c$ in Eq. \ref{eq:3} and off-diagonal terms cannot be assumed null because the reference frame is already being chosen such that $\Vec{B}_0$ lies on the $xz$ plane and the NV axis is along $z$. Therefore, the initial goal is to determine the hyperfine interaction matrix $A^c$ between an NV and a coupled $^{13}$C nucleus, such that the numerical simulations model can be validated with experimental data and subsequently extended to an arbitrary $^{13}$C coupling or even other spins.

\begin{figure}[t!]
	\includegraphics[width=\columnwidth]{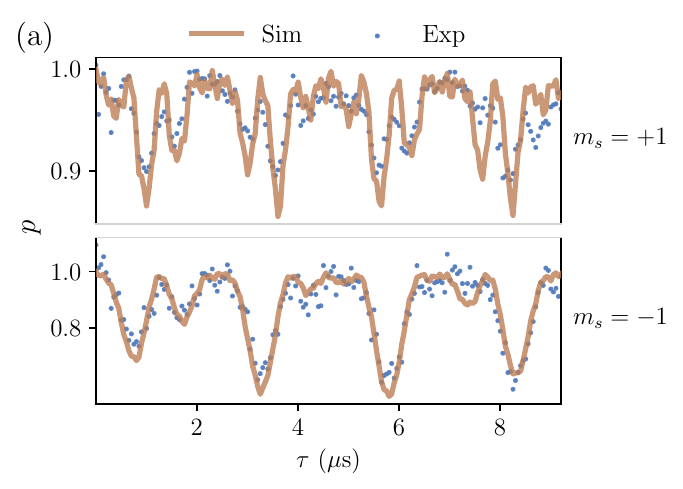}
	\includegraphics[width=\columnwidth]{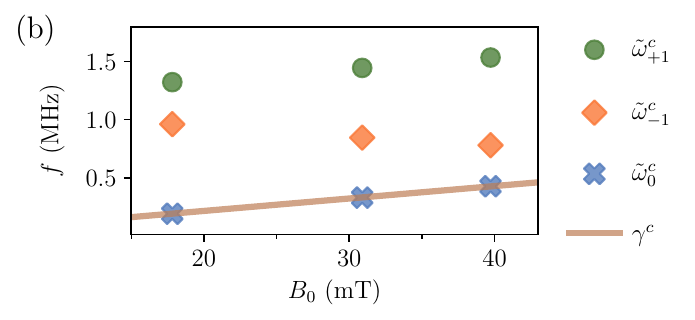}
	\caption{(a) ESEEM measured with the Hahn echo sequence for both $m_s=0 \rightarrow \pm 1$ transitions of a $^{14}$NV coupled to a $^{13}$C nuclear spin at $B_0=40$~mT. Both transitions show a slow modulation from the $\tilde{\omega}_{0}^c$ nuclear Larmor frequency at $m_s=0$ and fast modulations $\tilde{\omega}_{\pm 1}^c$ from the $m_s=\pm 1$ states. Correlations between experimental and simulated data are $r= $(0.761, 0.926). As in Fig. \ref{fig:15N} (a), the experimental fluorescence linearly fitted with the simulated transition probability can reach values above $p>1$ due to experimental noise. (b) Modulation frequencies obtained from the fit of the Hahn echo decay measurements as function of $B_0$. $\tilde{\omega}_{0}^c$ is closely related to the bare gyromagnetic ratio of $^{13}$C (solid line) due to an almost zero hyperfine interaction at $m_s=0$, while $\tilde{\omega}_{\pm 1}^c$ also decreases or increases by the same amount with an offset given by Eq. \ref{eq:6}. Error bars represented by the fit covariance are smaller than the points. Based on the experimental values of these modulations, the hyperfine tensor $A^c$ was determined by comparing them with the calculated eigenvalues from the total Hamiltonian.}
	\label{fig:13C_hahn}
\end{figure}

\begin{figure}[b!]
	\includegraphics[width=\columnwidth]{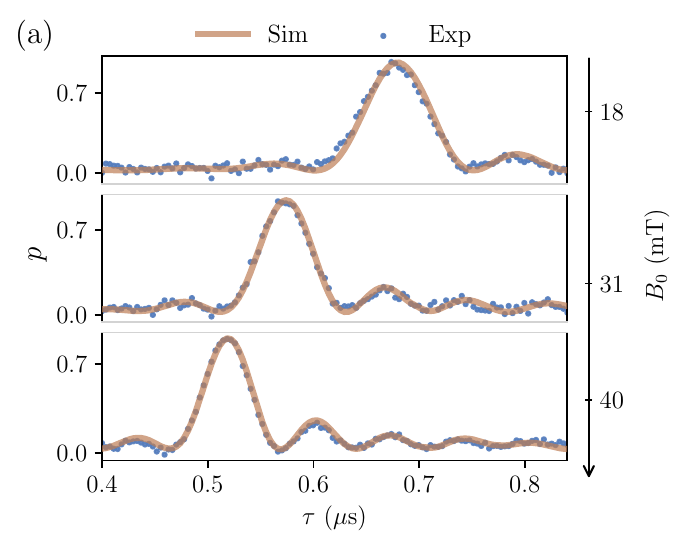}
	\includegraphics[width=\columnwidth]{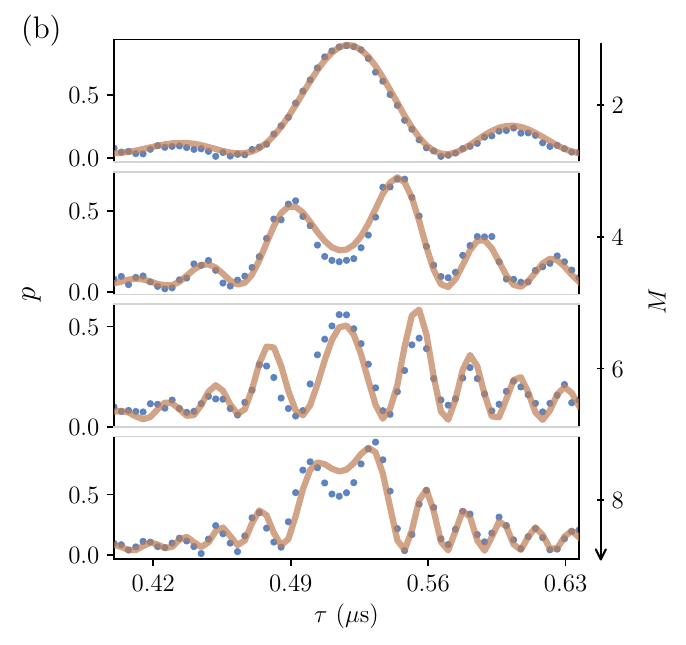}
	\caption{(a) XY8-2 resonances from a $^{14}$NV coupled to a $^{13}$C nuclear spin as a function of $B_0$ for the $m_s=0 \rightarrow +1$ transition. The calculated hyperfine matrix $A^c$ was used to simulate the resonances, with correlations $r=$ (0.987, 0.990, 0.989). As the magnetic field increases, the resonance shifts to larger frequencies, influenced by the positive gyromagnetic ratio of $^{13}$C. (b) Simulated and experimental XY8-$M$ at $B_0= 40$~mT, with correlations $r=$ (0.989, 0.977, 0.902, 0.951). As the order of the sequence increases, sidebands become more pronounced and line-widths smaller due to changes in the filter function. For this $^{13}$C coupling and these fields, these ambiguous resonances do not correspond to other nuclear spin. The high correlation between simulations and experiments attests the robustness of the numerical model also on higher dimensional Hilbert spaces, $\mathcal{H}_S \otimes \mathcal{H}_{I^n} \otimes \mathcal{H}_{I^c}$.}
	\label{fig:13C_XY8}
\end{figure}

Similar to the coupling with $^{15}$N, the Hahn echo decay of an NV coupled to a $^{13}$C nuclear spin is also expected to show periodic oscillations given by the effective Larmor frequencies of the nuclei $\tilde{\omega}_{m_s}^c$ \cite{13C_1}. Thus, Hahn echo was measured in a single $^{14}$NV for both electron spin transitions $m_s=0 \rightarrow \pm 1$ at three different magnetic fields, to avoid nitrogen modulations as in the previous section. The experimentally measured and later simulated modulations at $B_0 = 40$~mT are shown in Fig. \ref{fig:13C_hahn} (a), where no coherence decay can be observed at this time scale. The ESEEM for the two other values of $B_0$ can be found in Appendix \ref{S4:13C_hahn}. As with $^{15}$N coupling, the Hahn echo decay for $m_s=0 \rightarrow -1$ transition will present a slow modulation from $\tilde{\omega}_{0}^c$ and a fast one from $\tilde{\omega}_{-1}^c$, similarly to the $m_s=0 \rightarrow +1$ transition with $\tilde{\omega}_{+1}^c$ begin the fast modulation. The values of the nuclear Larmor frequencies $\tilde{\omega}_{m_s}^c$ obtained from the fit of the curve \cite{single3} are plotted against the bias field in Fig. \ref{fig:13C_hahn} (b). The $\tilde{\omega}_{0}^c$ frequency is closely related to the bare gyromagnetic ratio of $^{13}$C $\gamma^c = 10.705$ MHz/T, because of an almost zero hyperfine interaction at $m_s=0$. On the other hand, $\tilde{\omega}_{+1}^c$ and $\tilde{\omega}_{-1}^c$ respectively increases or decreases by the same amount based on the sign of $m_s$. At $B_0 = 0$ mT, both of them are given by \cite{13C_4}
\begin{equation}\label{eq:6}
	\tilde{\omega}_{\pm 1}^c (B_0=0)= \sqrt{(A_{zx}^c)^2 + (A_{zy}^c)^2 + (A_{zz}^c)^2}.
\end{equation}
Finally, several different combinations of the hyperfine elements in Eq. \ref{eq:3} were considered and the eigenvalues of the total Hamiltonian $H_0 + H_2$ calculated numerically. The hyperfine interaction matrix $A^c$ with the least square difference from the experimental values of $\tilde{\omega}_{m_s}^c$ is shown in Appendix \ref{S4:13C_hahn}. Using the obtained $A^c$, the Hahn echo decay was simulated [Fig. \ref{fig:13C_hahn} (a)], resulting in the same experimental modulations with high correlations of $r = 0.761$ for $m_s=+1$ and $r = 0.926$ for $m_s=-1$.

Based on the obtained hyperfine interaction matrix, XY8-2 measurement for the $m_s=0 \rightarrow +1$ transition was simulated and measured at the same bias magnetic fields, as shown in Fig. \ref{fig:13C_XY8} (a). The numerical simulation model and the calculated $A^c$ accurately describe the resonances induced by the $^{13}$C nuclear spin. As $B_0$ increases, the resonance shifts to shorter pulse separations and faster frequencies. This, in contrast to the coupling with $^{15}$N [Fig. \ref{fig:15N} (c)], evidences that the Zeeman interaction of the $^{13}$C nuclear spin with the external magnetic field is comparable in strength with the hyperfine interaction with the NV. Lastly, Fig. \ref{fig:13C_XY8} (b) presents XY8-$M$ measured and simulated at $B_0 = 40$~mT. Analogous to the $^{15}$N resonance [Fig. \ref{fig:15N} (d)], the linewidth becomes narrower and sidebands become more pronounced as the order of the sequence increases, which can be understood in terms of an alteration of the frequency filter function induced by the DD sequence \cite{phd_muller}. As the order increases, the correlation between experimental and simulation data decreases, possibly due to some pulse-length error [Appendix \ref{S5:pulse_errors}] or inaccuracy in the fitted ESEEM frequencies.

At the magnetic fields shown here, this particular NV exhibits $^{13}$C resonances which do not overlap with other possible nuclear spin species. However, for different couplings and fields, it may be the case that the interaction with $^{13}$C leads to the appearance of ambiguous resonances mimicking other nuclear species. Given the crystalline periodicity of the host diamond, a $^{13}$C atom can only occupy a limited number of lattice sites around the NV, resulting in a discrete set of possible hyperfine coupling strengths divided into families \cite{13C_2, 13C_3, 13C_4}. Thus, the dataset provides simulations for hyperfine parameter values calculated from density functional theory \cite{13C_4}, such that a few possible $^{13}$C resonances can be disambiguated from experimental data at different fields. A more conclusive method to identify the nuclear species giving rise to some resonance is, however, to measure ESEEM frequencies and compare the slow modulation with the expected gyromagnetic ratio, as performed here. Although we focused the discussion on lattice $^{13}$C nuclear spins, the same method is applicable to an arbitrary spin coupled to the NV, as long as their interaction can be modeled by Eq. \ref{eq:3}.

	
\subsection{Finite Pulse Duration} \label{sec:pulse_duration}

In most applications of multipulse sequences, perfect $\delta$-pulses with zero temporal length are assumed. Even though the pulse length might still be much smaller than the pulse separation $t_\pi \ll \tau$, it has been shown that the free evolution of the system during finite pulse durations under the influence of $H_0$ leads to additional resonances at even multiples of the resonant frequencies and their odd subharmonics \cite{spurious}, that is $\tau=(2k+1)/(2l) \tau_0$ for $k$ and $l$ natural. Notably, lattice $^{13}$C nuclear spins can give rise to ambiguous resonances of $^1$H, other than the ones already described in the previous section. This occurs because the 4\textsuperscript{th} harmonic of a weakly coupled $^{13}$C coincides with the fundamental resonance of $^1$H, given that $4 \gamma^c = 42.82$~MHz/T and $\gamma^h=42.57$ MHz/T. These ambiguous resonances can still be greatly suppressed by adding a random phase $\Phi_m$ for each XY8$_m$ block, known as the RXY8-$M$ sequence \cite{RXY8_1, RXY8_2}. A further improvement of this suppression was theoretically proposed by correlating the random phases such that they cancel each other every $g$ repetitions of the XY8 block, $\sum_{m}^{g} \exp(-i \Phi_m) = 0$ \cite{RXY8_g}.

\begin{figure}[t!]
	\includegraphics[width=0.49\textwidth]{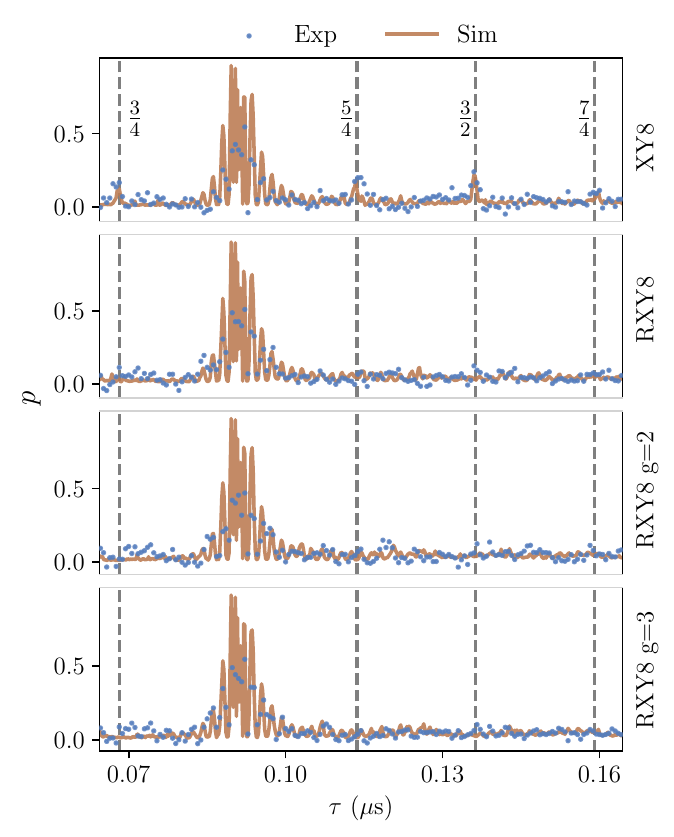}
	\caption{XY8-12 and RXY8-12 sequences with resonance from an external RF signal with $f_0=5.5$ MHz for $m_s=0\leftrightarrow-1$ transition. Multiple fringes are observed around the fundamental resonance $\tau_0=1/2f_0$, in addition, spurious harmonics due to the finite pulse length at $\tau/\tau_0 =$ 3/4, 5/4, 3/2 and 7/4 are suppressed by the phase randomization \cite{RXY8_1, RXY8_2}. A proposed improvement of the RXY8 sequence by random phase correlations with $g=2$ and 3 \cite{RXY8_g} is also demonstrated, without significant differences from uncorrelated RXY8. Thus, indicating that phase correlation can be introduced in the RXY8 sequences, without affecting non-ambiguous spectral features. The best correlations $r=$(0.839, 0.877, 0.818, 0.888) occur at $\gamma_e B_{2}^z = 0.28$ MHz and $\gamma_e B_{2}^x = \gamma_e B_{2}^y = 0$ from Eq. \ref{eq:4}.}
	\label{fig:RXY8_g}
\end{figure}

To experimentally reproduce these ambiguous resonances, an external RF signal with frequency $f_0 = 5.5$~MHz was applied to a single NV at bias field $B_0=38$~mT, where the MW $\pi$-pulse length was $t_\pi=17.16$~ns. Four different multipulse sequences were applied as shown in Fig. \ref{fig:RXY8_g}. In all four we observe multiple fringes around the fundamental resonance $\tau_0=1/(2f_0)$, resulting from the convolution of the spectral density of the RF field and the filter function of the multipulse sequence \cite{phd_muller}. Apart from that, XY8-12 has four of these ambiguous resonances at $\tau/\tau_0 =$ 3/4, 5/4, 3/2 and 7/4, which are almost entirely suppressed with RXY8-12. Furthermore, we demonstrate the experimental realization of RXY8-12 with correlated phases at $g=2$ and $g=3$ repetitions. At these experimental conditions, the two sequences show no noticeable differences from uncorrelated RXY8. Meaning that they can be used instead of the RXY8, as they have additional robustness against pulse errors \cite{RXY8_g}, such as in the pulse length and detuning of the MW resonant frequency. These lead to a degradation of the resonance contrasts and are discussed in more details in Appendix \ref{S5:pulse_errors}.

Eq. \ref{eq:4} was used to model the RF interaction with the NV, with the best correlations found for $\gamma_e B_{2}^z = 0.28$~MHz and  $\gamma_e B_{2}^x = \gamma_e B_{2}^y = 0$. The simulations also show a suppression of the spurious harmonics by the phase randomization. Hence, demonstrating that the numerical simulations model can be applied for time-dependent sensing Hamiltonians $H_2(t)$, while also taking into account the specific time dependency of the control Hamiltonian $H_1(t)$ from Eq. \ref{eq:2}.


\section{Discussion and Simulations Dataset} \label{sec:discussion}

Noise filtering by DD sequences has proven to be an effective technique in quantum sensing, capable of detecting very weak oscillating fields of even single nuclear spins. This great sensitivity comes at the cost of a complex signal analysis influenced by external and inherent interactions which lead to ambiguous resonances, such as the ones studied here. It is therefore imperative that these effects are well understood and characterized for an accurate identification of nuclear spin resonances. Among candidate platforms, the NV center in diamonds outstands in its ease of implementation and versatility with different biological and chemical systems. Nevertheless, widely adopted approximations and oversimplifications of the system dynamics completely overlook these ambiguous resonances. Examples of such approximations are the RWA, assuming perfect $\delta$-pulses, adopting a reduced electronic spin manifold where $S=1/2$ or ignoring the nitrogen nuclear spin of the NV.

In this work, we presented a numerical simulations model for the time-dependent dynamics of the NV under a DD sequence and experimentally characterized three effects that lead to ambiguous resonances. First, in $^{15}$NV centers, a small misalignment of $\Vec{B}_0$ causes the precession of $^{15}$N nuclear spins to be sensed via the hyperfine interaction. As demonstrated here, these resonances can closely resemble a $^{1}$H nuclear spin. Another relevant source of ambiguous resonances are the diamond lattice $^{13}$C nuclear spins, which also couple to the NV electronic spin through the hyperfine interaction matrix $A^c$. Here, we have presented a method for unequivocally calculating $A^c$, based on the ESEEM frequencies as a function of the bias field $B_0$. This method can be generalized to any coupled spins in the strong coupling regime \cite{strong_coupling}, which can be expressed in terms of a Hamiltonian as in Eq. \ref{eq:3}. Lastly, specific parameters of the MW control pulses which are not accounted for in perturbation theory also affect the resonances of DD sequences. In particular, the free evolution of the probing spin during finite pulse durations leads to resonances at the even multiples of the fundamental resonant frequencies \cite{spurious}. These resonances can be suppressed with phase randomization in each XY8 block \cite{RXY8_1, RXY8_2}, where we experimentally demonstrated a proposed improvement with random phase correlation \cite{RXY8_g}.

Unlike the ambiguous resonances originating from the finite pulse durations, a straightforward method for suppressing the resonances from $^{15}$N and $^{13}$C remains an problem. Currently, this would involve more elaborate experimental conditions, as precise vector magnets, isotopically purified diamonds or more complicate pulse sequences. Another important technique that can be potentially employed for improving pulse fidelity and compensate for finite-length hard pulses is pulse shaping \cite{pulse_shaping}, which could be easily incorporated in our numerical simulations model. In conclusion, some of the ambiguous resonances presented here are simply unavoidable in a large variety of experiments. A platform for analyzing and identifying spectral features is thus essential for the continued development of the field, as these ambiguous resonances are largely dependent on specific experimental parameters, resulting in numerous possible resonant responses in multipulse DD sequences.

To further extend the toolbox of available methods, we provide a dataset comprised of around $10^4$ simulations of XY8-$M$ for $^{15}$N and $^{13}$C ambiguous resonances. Both $m_s$ transitions were considered with several different combinations of $B_0$ and order $M$. It is important to note however, that a fixed Rabi frequency of $\gamma_e B_1 = 40$~MHz was considered in all simulations of the dataset, which may lead to slight differences at high order XY8. This typically high Rabi frequency was chosen in order to mitigate the ambiguous resonances from finite pulse durations and thus not overlap it with the two other sources of ambiguous resonances. For $^{15}$N resonances, the angle $\theta_0$ was changed, whereas for $^{13}$C, the hyperfine matrices of all provided families \cite{13C_4} were considered. The simulations can be visualized and compared with the user's experimental data for spectral disambiguation through an open source and user-friendly graphical interface, published under \href{https://figshare.com/articles/dataset/Dataset_for_Ambiguous_Resonances_in_Multipulse_Quantum_Sensing_with_NVs/26245895}{DOI:10.6084/m9.figshare.26245895} \cite{dataset}, where a complete guide for installation and usage are also provided. Even though the dataset only contains XY8-$M$ simulations for NVs, it can be used for other sequences which have similar resonance spectra, such as the CPMG \cite{CPMG}.

In the ambiguous resonances we studied here, the simulations model presented robust correlations with the experimental data. Numerically solving the system's time dependent dynamics without the RWA has proven to be a more versatile and robust approach than semi-analytical models, recurring to perturbation theory. Although primarily focused with NVs, the developed model can be applied to any quantum system which satisfies the following three criteria - it can be described in terms of the density matrix formalism, it has a total Hamiltonian which can be divided in three parts as $H = H_0 + H_1(t) + H_2(t)$ and it has a hermitian observable. Most color centers and many other quantum systems satisfy these conditions. Furthermore, the generality of $H_1(t)$ implies that the model can be extended to more elaborate DD sensing protocols as well.
	
	\begin{acknowledgments}
		
		We acknowledge Miriam Mendoza Delgado and Prof. Dr. Cyril Popov from the Institute of Nanostructure Technologies and Analytics of the University of Kassel for the annealing a diamond sample. We are also grateful to Karolina Schüle, Jens Fuhrmann and Prof. Dr. Fedor Jelezko from the Institute for Quantum Optics of Ulm University for providing another diamond sample. Finally, we would like to thank Alexander Külberg and Dr. Andreas Thies from Ferdinand-Braun-Institute (FBH) for performing the ion implantations. This work was supported by the German Federal Ministry of Education and Research (BMBF) under the Grand Challenge of Quantum Communication, in a collaboration between "\textit{DIamant-basiert QuantenTOKen}" (DIQTOK - n\textsuperscript{o} 16KISQ034) and  “Quantum Photonic Integrated Scalable memory" (QPIS - n\textsuperscript{o} 16KISQ032K) projects. In addition, this work received funding from BMBF project DiNOQuant (No. 13N14921) and German Research Foundation (DFG) grants 410866378 and 41086656.
		
	\end{acknowledgments}
	
	\section*{Data \& Code Availability}
	
	All data and code used in this work are available for scientific use upon reasonable request.
	

\appendix

\section{Experimental Setup and Data Acquisition} \label{S1:exp}

The experiments were performed on a custom-made confocal microscope. The diamond samples are mounted on a custom made 3D stage movable with 10 $\mu$m precision. A diode laser with a wavelength of 518~nm (iBEAM-SMART-515-S, TOPTICA Photonics AG) was used for both continuous and pulsed optical excitation of an NV center. Emitted fluorescence was collected either with an air objective (MPlanApo N 50$\times$/0.95, Olympus) with NA~=~0.95 or an oil objective (HCX PL APO 100$\times$/1.40, Leica; type F immersion liquid, Leica) with NA~=~1.40 and detected by an avalanche photodiode (SPCM-AQRH-14, Excelitas). The objective is screwed on a piezo scanner (P-517, Physikinstrumente) with subnanometer resolution. The laser light was blocked from entering the detector by a 650 nm long pass filter (FELH0650, Thorlabs). Continuous MW field for ODMR spectra was generated by a microwave source (TGR6000, TTi) and the RF signal in Sec. \ref{sec:pulse_duration} by a function generator (TGF4242 LXI, TTi). MW pulses for dynamical decoupling multipulse sequences were generated by an arbitrary waveform generator (AWG7122C, Tektronix), with maximum sampling frequency of 12~GSamples/s and memory of 64~MSamples. In both cases, generated MW fields were amplified (ZHL-16W-43+, MiniCircuit) and combined with the non-amplified RF using a bias tee (Mini Circuits, ZABT-2R15G+). Finally, they were transmitted to the NV centers by a copper (99.99\%) wire with a diameter of 20~$\mu$m. The bias magnetic field was applied with a permanent magnet mounted on a holder with 4 degrees of freedom and with the maximum achievable magnetic field projection on NV center axis of 0.1~T. For time-resolved photon counting we used a time-correlated single photon counter (MCS6, FAST ComTec). Prior to all pulsed measurements, the $\pi$-pulse duration was measured with a Rabi sequence. The experimental value of the Rabi frequency was then used in the simulations. Lastly, laser pulses for state polarization and readout were applied for 3~$\mu$s and power from 0.1 to 0.5~mW, depending on the sample.


\section{Diamond Samples} \label{S2:diamonds}

In Sec. \ref{sec:15N}, the effects of $^{15}$N coupling on Hahn echo were measured in a IIa diamond with a $^{12}$C overgrown layer of 90 nm, which was later implanted with $^{15}$N at energy 2~keV and annealed at 1000$^\circ$C in ultra-high vacuum. For the XY8-$M$ measurements, another shallow implanted sample was used with implantation energy 2 keV and dose $10^{9}$~$^{15}$N$^+$/cm$^2$, annealed at 10$^{-7}$~mbar and 1000$^\circ$C. In Sec. \ref{sec:13C}, the $^{13}$C ambiguous resonance were studied in an electronic grade diamond grown by chemical vapor deposition, with natural abundance of $^{13}$C and $^{14}$N, in order to avoid $^{15}$N effects. Finally, the ambiguous resonances from finite pulses in Sec. \ref{sec:pulse_duration} were measured in a thin electronic grade diamond plate with deeper $^{15}$N implantation (20-30 keV) and doses around $10^9$~$^{15}$N$^+$/cm$^2$, also annealed in ultra high vacuum at 1000$^\circ$C.


\section{Pearson Correlation Coefficient} \label{S3:pearson}

\begin{figure}[t!]
	\includegraphics[width=.45\textwidth]{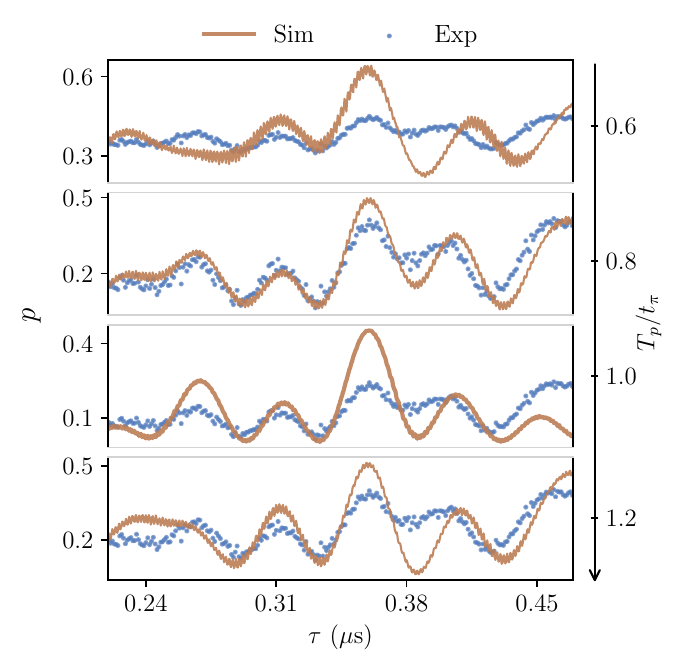}
	\caption{The XY8-2 of the ambiguous $^{15}$N resonance (Sec. \ref{sec:15N}) as a function of the pulse duration $T_p/t_\pi$ exemplifies the use of the Pearson coefficient, with r = (0.419, 0.858, 0.555, 0.685) for $T_p/t_\pi=$ (0.620, 0.826, 1.000, 1.215) respectively. At $T_p/t_\pi=1$ the simulated and experimental data have a poor correlation, but considering a small deviation, the experimental data can be better modeled by the simulation. The pulse error induces a high frequency modulation of the signal and variations in the peak intensities.}
	\label{fig:t_pi}
\end{figure}

As discussed in Sec. \ref{sec:methods}, experimental fluorescence intensity and simulated transition probability to the bright state ($m_s=0$) are related through a linear equation. Therefore, an appropriate metric to determine the correlation between both is the Pearson coefficient $r$ \cite{pearson}, which is the square root of the coefficient of determination $R^2$ for linear relations. It is defined as the ratio between the covariance of the two data sets and the product of their standard deviations, where one data set represents the experimental data and the other the simulation. Within this definition, $r$ is normalized from -1 to 1, with $r=0$ representing no linear dependency between experimental and simulated data, while $r=1$ indicates a perfect linear correlation. 

To illustrate this, Fig. \ref{fig:t_pi} shows simulations for XY8-2 of the $^{15}$N ambiguous resonances at $B_0=17$ mT for different pulse length deviations in regards to the actual $\pi$-pulse duration $T_p/t_\pi$ (Sec. \ref{sec:15N}). Compared to the actual experimental data, assuming $T_p/t_\pi=1$ results in a low correlation of $r=0.555$ and thus an inadequate simulation of the experimental data, whereas taking $T_p/t_\pi=0.826$ provides a better correlation of $r=0.858$. In this case, the pulse length deviation affects the relative intensity of each resonance peak keeping their $\tau$ fixed, while also inducing a high frequency modulation of the signal.


\section{\textsuperscript{13}C Hahn Echo Modulations} \label{S4:13C_hahn}

\begin{figure*}[t!]
	\includegraphics[width=0.85\textwidth]{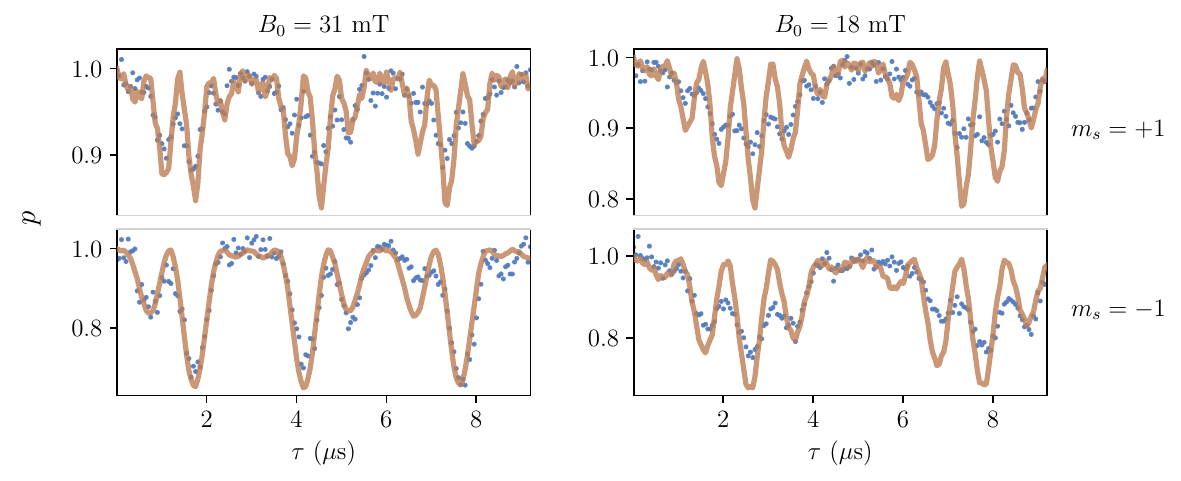}
	\caption{Hahn echo modulations from a $^{14}$NV coupled to a $^{13}$C nuclear spin, resulting in correlations between experimental and simulated data of $r=$ (0.792, 0.925) at $B_0=31$ mT and $r=$ (0.659, 0.806) at $B_0=18$ mT. Although both agree in regards to the modulation frequencies, the amplitude of the faster modulation is smaller in the experimental data, which could be due to some experimental error or unconsidered factor in the Hamiltonian.}
	\label{fig:13C_hahn_B0}
\end{figure*}

The ESEEM measured by a Hahn echo sequence at $B_0 =31$ mT and $B_0=18$ mT for both $m_s$ transitions are shown in Fig. \ref{fig:13C_hahn_B0}. As can be observed, the simulations and experimental data both present the same modulation frequencies, however the amplitude of the fast modulation is smaller in the experiments. This could be related to some experimental error or some unconsidered factor in the Hamiltonian model. Nonetheless, this discrepancy does not affect the study of the ambiguous resonances and as seen in Sec. \ref{sec:13C}, the calculated hyperfine matrix and the numerical model still reliably described them.

Based on those experimental values of the modulation frequencies, different combinations of hyperfine coupling matrices between the NV and the $^{13}$C nuclear spin were tested in the total Hamiltonian $H_0 + H_2$ from Eqs. 1 and 3. The eigenenergies of the Hamiltonians were calculated numerically, with the effective Larmor frequencies $\tilde{\omega}^c_{m_s}$ corresponding to the energy differences between the carbon nuclear spin states in each electron spin state $m_s$. Since no splittings could be observed in the two $m_s$ resonances, the ODMR spectrum sensitivity sets an upper bound for the $|A_{zz}^c|$ component of 4 MHz. Furthermore, the non-diagonal components of the $^{13}$C families are lower or of the same order as $|A_{zz}^c|$ \cite{13C_4}, thus this bound was also assumed for the remaining components of $A^c$. Initially then, the six hyperfine elements were linearly varied in the interval from -4.0~MHz to 4.0~MHz with a large spacing of 0.5~MHz. Having found an approximate optimal region of the parameter space, each parameter was subsequently varied with spacing 0.01~MHz in this smaller range, totaling over $10^7$ possible combinations. Ultimately, the hyperfine matrix with the least square difference in units of MHz was found to be:
\begin{equation*}
	A^c = \begin{pmatrix}
		-0.25 & -1.85 & -0.49 \\
		-1.85 & 0.00 & 0.01 \\
		-0.49 & 0.01 & 1.01
	\end{pmatrix} .
\end{equation*}


\section{Pulse Errors} \label{S5:pulse_errors}

\begin{figure}[b!]
	\includegraphics[width=0.49\textwidth]{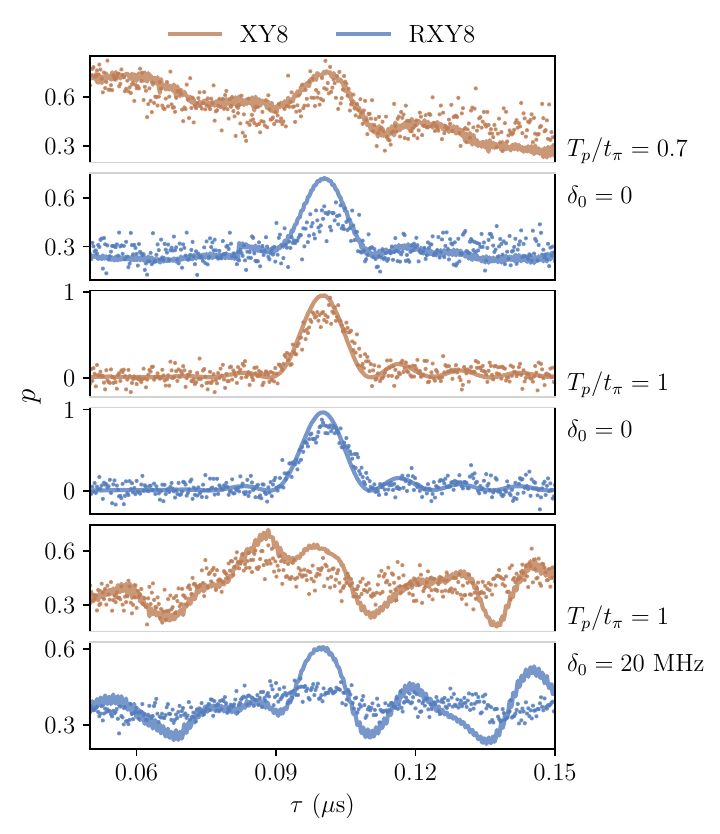}
	\caption{XY8-2 and RXY8-2 with external RF signal of $f_0=5$ MHz with MW pulse deviation of $\delta_0=20$ MHz or length error of $T_p/t_\pi=0.7$. The RF field is approximated by an effective hyperfine coupling with $A^s_{xz} = 0.92$ MHz, resulting in $r=$(0.763, 0.907, 0.626) for XY8-2 and $r=$(0.639, 0.915, 0.423) in RXY8-2. The pulse errors lead to a degradation of the contrast and additional resonances, which are suppressed by the random phases.}
	\label{fig:RXY8_errors}
\end{figure}

Phase randomization can greatly increase robustness against pulse errors \cite{RXY8_1, RXY8_g}. In non-randomized XY8, small detunings in the MW frequency $\delta_0$ and deviations in the pulse length $T_p/t_\pi$ lead to a degradation of the resonance contrast and additional peaks over the $\tau$ range, which could represent another source of an ambiguous resonance for similar frequency RF fields or spins. To show this effect, XY8-2 and RXY8-2 with uncorrelated phases were measured and simulated with an applied RF field of $f_0 = 5$ MHz, taking $\delta_0=20$ MHz and $T_p/t_\pi=0.7$ separately (Fig. \ref{fig:RXY8_errors}). Experimental parameters are $t_\pi=15.24$ ns and $B_0=34$ mT with $\theta_0=4.1^\circ$. Again, we observe a suppression of the spurious peaks in RXY8-2 and a further validation of the simulation model. In this case at low order $M=2$, the RF signal can be approximated by an effective hyperfine interaction with $A^s_{xz} = 0.92$ MHz and $\gamma^s B_0 = 5$ MHz, as in Eq. \ref{eq:4}. Whereas for the higher order XY8-12 from Sec. \ref{sec:pulse_duration}, the quantum mechanical approximation of the classical RF field already deviates from the experimental data. At the large detuning considered here $\delta_0 = 20$~MHz, the simulation shows smaller correlation with the experimental data. This evidences a limitation of the numerical model, while the reasons for it remain an open problem. Nonetheless, such large detuning might not be of practical significance for most experiments, apart from benchmarking the numerical simulations model.


	
\end{document}